\documentclass[sn-nature,iicol]{sn-jnl}% Default with double column layout
\usepackage{graphicx}%
\usepackage{multirow}%
\usepackage{amsmath,amssymb,amsfonts}%
\usepackage{amsthm}%
\usepackage{mathrsfs}%
\usepackage[title]{appendix}%
\usepackage[table]{xcolor}%
\usepackage{textcomp}%
\usepackage{manyfoot}%
\usepackage{booktabs}%
\usepackage{algorithm}%
\usepackage{algorithmicx}%
\usepackage{algpseudocode}%
\usepackage{listings}%
\usepackage{placeins}
\usepackage{arydshln}
\usepackage{float}
\usepackage{tabularx}
\usepackage{soul}
\begin{document}
%%%%%%%%%%%%%%%%%%%%%%%%%%%%%%%%%%%%%%%%%%%%%%%%%%%%%%%%%%%%%%%%%%%%%%%%
\title [Artificial Intelligence for Food Innovation]
%{{\sffamily{\bfseries{{\hspace*{3.9cm}} Got meat?\\
{\sffamily{\bfseries{
%\hspace*{1.8cm} 
Artificial Intelligence for Food Innovation}}} 
%%%%%%%%%%%%%%%%%%%%%%%%%%%%%%%%%%%%%%%%%%%%%%%%%%%%%%%%%%%%%%%%%%%%%%%%
\author[]{\fnm{Bianca}    \sur{Datta}}      
%\email{biancad@gfi.org}
\author[]{\fnm{Markus J.} \sur{Buehler}}    
%\email{mbuehler@mit.edu}
\author[]{\fnm{Yvonne}    \sur{Chow}}       
%\email{yvonne\_chow@a-star.edu.sg}
\author[]{\fnm{Kristina}  \sur{Gligori\'{c}}} 
%\email{gligoric@jhu.edu}
\author[]{\fnm{Dan}       \sur{Jurafsky }}  
%\email{jurafsky@stanford.edu}
\author[]{\fnm{David L.}  \sur{Kaplan}}     
%\email{david.kaplan@tufts.edu}
\author[]{\fnm{Rodrigo}   \sur{Ledesma-Amaro}} 
%\email{r.ledesma-amaro@imperial.ac.uk}
\author[]{\fnm{Giorgia}   \sur{Del~Missier}}
%\email{giorgia.del-missier22@imperial.ac.uk}
\author[]{\fnm{Lisa}      \sur{Neidhardt}}  
%\email{ln327@cam.ac.uk}
\author[]{\fnm{Karim}     \sur{Pichara}}    
%\email{karim@thenotcompany.com}
\author[]{\fnm{Benjamin}  \sur{Sanchez-Lengeling}} 
%\email{ben.sanchez@utoronto.ca}
\author[]{\fnm{Miek}      \sur{Schlangen}}  
%\email{mies@igt.sdu.dk}
\author[]{\fnm{Skyler R.} \sur{St.~Pierre}} 
%\email{sstpie@stanford.edu}
\author[]{\fnm{Ilias}     \sur{Tagkopoulos}} 
%\email{itagkopoulos@ucdavis.edu}
\author[]{\fnm{Anna}      \sur{Thomas}}     
%\email{thomasat@stanford.edu}
\author[]{\fnm{Nik}       \sur{Watson}}     
%\email{n.j.watson@leeds.ac.uk}
\author[]{\fnm{Ellen}     \sur{Kuhl}}       
\email{ekuhl@stanford.edu}
%%%%%%%%%%%%%%%%%%%%%%%%%%%%%%%%%%%%%%%%%%%%%%%%%%%%%%%%%%%%%%%%%%%%%%%%
\abstract{Global food systems must deliver nutritious, sustainable foods 
while sharply reducing environmental impact. 
Yet, food innovation remains 
slow, empirical, and fragmented. 
Artificial intelligence (AI) offers a transformative path
to link molecular composition to functional performance, 
connect chemical structure to sensory outcomes, and 
accelerate cross-disciplinary innovation
across the production pipeline. 
While broadly applicable to food systems, 
we focus on sustainable proteins--plant-based, fermentation-derived, and cultivated--as a high-impact testbed for AI-driven closed-loop design.
We review 
the applications, opportunities, and challenges 
of AI for Food 
as an emerging discipline that integrates 
ingredient design, 
formulation development, 
fermentation and production, 
texture analysis, 
sensory science,
manufacturing, and 
recipe generation.
We identify four priorities: 
advancing scientific machine learning with embedded domain priors,
treating food as a programmable biomaterial, 
building self-driving laboratories for automated discovery, and 
developing deep reasoning models that integrate nutrition and sustainability. 
Integrating AI responsibly into the food innovation cycle
can accelerate the transition to sustainable food systems 
and establish a predictive, design-driven science of food 
for human and planetary health.\\[10.pt]
\hspace*{6.0cm}{{\textbf{Editor's summary}}}\\[3.pt]
Artificial intelligence is transforming how new foods are designed, produced, and experienced. While AI has the potential to link molecular composition to functional and sensory performance and accelerate food innovation, challenges related to data quality, transparency, and governance remain. Addressing these requires interdisciplinary collaboration, open and inclusive data ecosystems, and AI systems that augment rather than replace human expertise.}
%%%%%%%%%%%%%%%%%%%%%%%%%%%%%%%%%%%%%%%%%%%%%%%%%%%%%%%%%%%%%%%%%%%%%%%%%%%%%
\maketitle
%%%%%%%%%%%%%%%%%%%%%%%%%%%%%%%%%%%%%%%%%%%%%%%%%%%%%%%%%%%%%%%%%%%%%%%%%%%%%
%%%%%%%%%%%%%%%%%%%%%%%%%%%%%%%%%%%%%%%%%%%%%%%%%%%%%%%%%%%%%%%%%%%%%%%%%%%%%
\begin{figure*}[h]
    \centering
   \includegraphics[width = 1.0\textwidth]{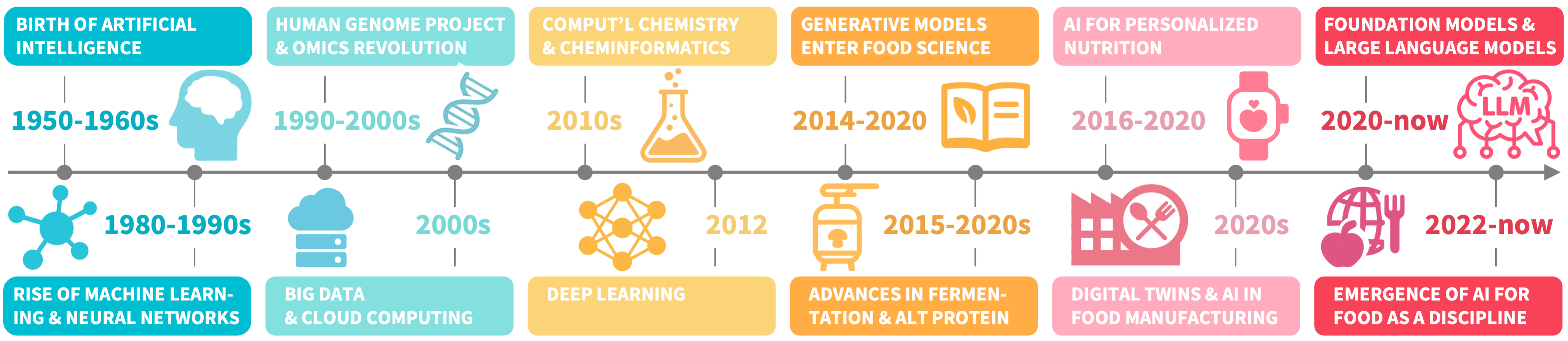}
    \caption{\sffamily{\bfseries{Historical milestones in AI for Food.}} 
The birth of AI in the 1950s–1960s, marked by Alan Turing's concepts of computing machinery and intelligence, established the foundations of machine intelligence. The rise of machine learning and neural networks in the 1980s–1990s, driven by backpropagation and statistical learning, shifted AI from rule-based systems to data-driven approaches. In parallel, the human genome project and the omics revolution in the 1990s–2000s, enabled by high-throughput sequencing, created a culture of large-scale biological and nutritional data. The expansion of big data and cloud computing in the 2000s further empowered the analysis of massive ingredient, supply chain, and consumer preference datasets.
Since the beginning of the 21st century, advances in computational chemistry and cheminformatics in the 2010s have enabled the prediction of flavor molecules, aroma profiles, and chemical interactions. Breakthroughs in deep learning, exemplified by AlexNet in 2012, catalyzed applications in food recognition and image-based dietary tracking. Since 2014, generative models have expanded into food design, flavor and recipe generation, and ingredient discovery, with companies such as NotCo applying AI to plant-based food development. Since 2015, advanced fermentation and alternative protein models have started to include AI-driven strain engineering and process optimization to accelerate precision fermentation and industrial adoption.
The current decade is witnessing the integration of wearable and microbiome data for AI-driven personalized nutrition since 2016, the emergence of digital twins for real-time monitoring and optimization of food manufacturing processes, and the rise of foundation models and large language models that leverage unstructured data for formulation and recipe innovation since the 2020s. Together, these advances are fueling the emergence of AI for Food as a distinct scientific discipline.}
\label{fig01}
\end{figure*}
%%%%%%%%%%%%%%%%%%%%%%%%%%%%%%%%%%%%%%%%%%%%%%%%%%%%%%%%%%%%%%%%%%%%%%%%%%%%%%
Food is among the most powerful determinants of human health, planetary sustainability, and cultural identity \cite{lappe71}; yet, research and development for food remains slow, fragmented, and driven by trial-and-error rather than rational design. Radical, urgent, and creative food solutions are needed to combat the current climate crisis \cite{clark20}. Notably, across farmland, livestock, and production, the food sector generates 35\% of all global greenhouse gas emissions  \cite{xu21}, and contributes heavily to biodiversity loss and public health challenges \cite{keesing22}. To meet climate goals and feed a growing population, we need alternative sources of protein that are delicious, affordable, scalable, and nutritionally compelling \cite{friedrich22}. Today, however, the scientific bottlenecks in texture, flavor, scalability, and cost pose barriers to adoption \cite{barabasi20}. Artificial intelligence (AI) has reshaped domains, from protein design and drug discovery to materials engineering \cite{zeni25}. AI is now poised to revitalize how we grow and make food--from optimizing plant-based meats to generating climate-conscious recipes \cite{alsarayreh23}. Enabling protein diversification will have profound implications, and AI can substantially accelerate scientific progress if applied responsibly and towards the correct challenges \cite{datta22}. But the success of AI is by no means guaranteed. When deployed in a scattershot way, AI can generate low-quality outputs, increase environmental burdens, or undermine consumer trust, aspects that are especially sensitive in food \cite{king25}. The goal is thus not to deploy AI indiscriminately. Instead, we need to partner with AI where it measurably reduces time, cost, and uncertainty for problems that matter \cite{kuhl25}. Responsible application--focused on the right challenges--is essential. In this review, we identify areas across the food system where we can constructively engage with AI: we provide background on the current state of the art, outline new opportunities, and discuss the challenges that AI is facing in the quest to accelerate the future of food. 

The intellectual and technological trajectory 
that has shaped to the emergence of {\it{AI for Food}} as a discipline 
spans more than seven decades, 
weaving together advances in AI, food science, and biotechnology (Fig. \ref{fig01}).
From the birth of symbolic AI in the 1950s \cite{turing50},
and the rise of machine learning and neural networks in the 1980s–90s \cite{rumelhart86}, 
to the deep learning breakthroughs in the 2010s \cite{krizhevsky17},
and the advent of generative models \cite{goodfellow14} and foundation models \cite{vaswani17} in the 2020s, 
AI has progressively expanded its ability to model, design, and predict complex systems. 
In parallel, developments in genomics \cite{ihgsc01}, 
computational chemistry \cite{gilmer17}, 
fermentation, and personalized nutrition \cite{zeevi17} 
have created unprecedented data resources and biological insights directly relevant to food design. 
Their convergence has given rise to a new field in which AI is now central to ingredient discovery, recipe formulation, manufacturing optimization, and consumer personalization \cite{kuhl25}. 

Designing food materials poses unique challenges, as their properties emerge from multiscale interactions among proteins, carbohydrates, lipids, and processing conditions \cite{gordon25}. Yet, AI is shifting food innovation from trial-and-error to data-driven discovery. Modern models link chemical composition to sensory outcomes and enable targeted formulation and rapid iteration. Machine-learning frameworks predict consumer preferences from chemical–sensory panels \cite{schreurs24}, while graph neural networks map molecular structure to odor quality \cite{lee23}. Advances in protein modeling and molecular learning accelerate functional ingredient discovery \cite{jumper21,butler18}, and precision fermentation provides scalable routes to novel proteins and textures \cite{graham23}. 
Together, these advances address long-standing challenges---engineering egg-free emulsions, meltable dairy alternatives, or fibrous plant-based meats---and have translated into market-ready products from companies such as NotCo, Perfect Day, and Climax Foods \cite{lurieluke24}. This transition from conceptual promise to industrial deployment underscores AI’s growing role as a driver of culinary innovation and sustainability.

\begin{figure}[t]
    \centering
   \includegraphics[width = 0.48\textwidth]{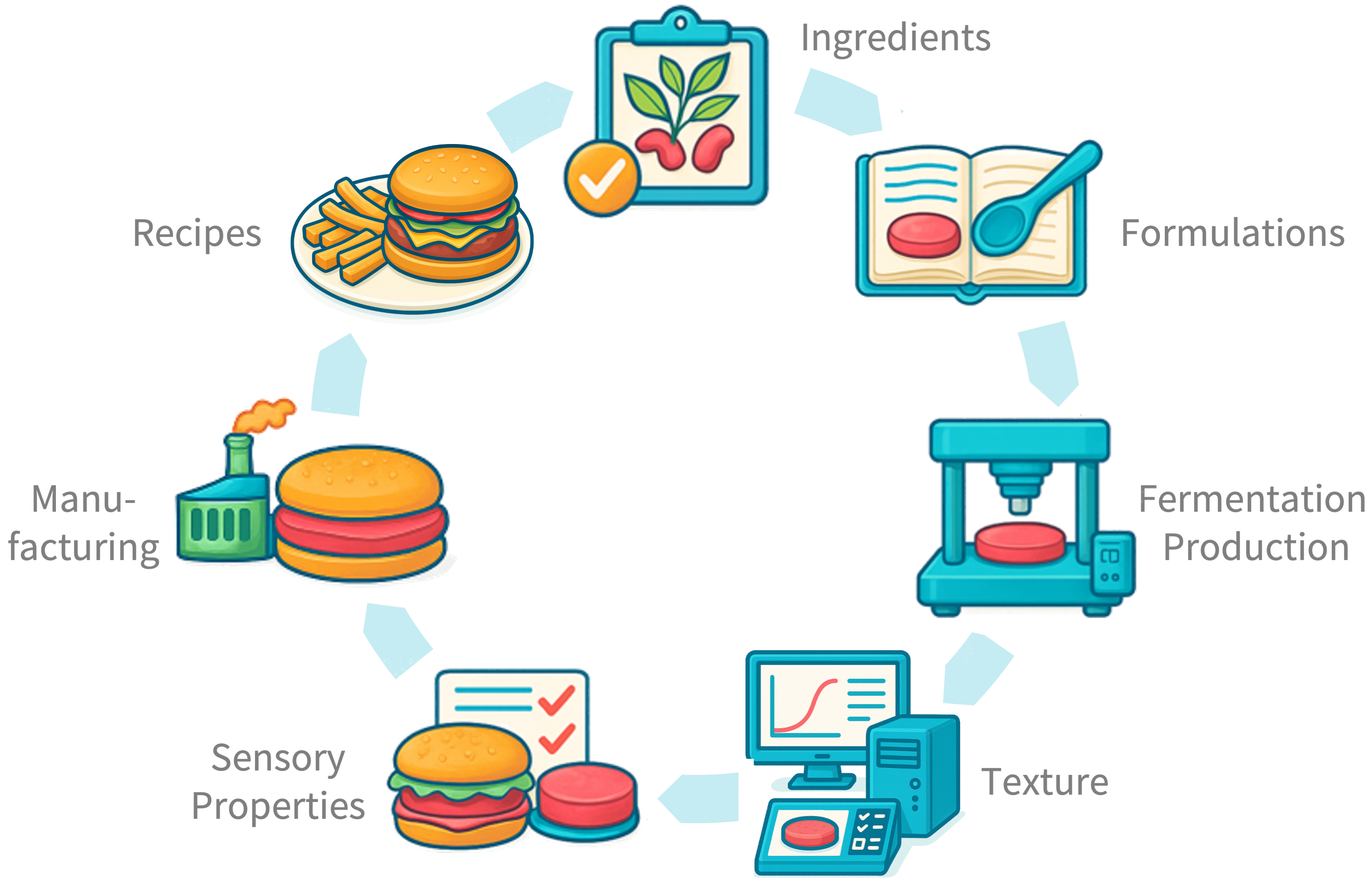}
    \caption{\sffamily{\bfseries{The AI-powered production cycle.}}    
Seven key stages where artificial intelligence can accelerate the time to market: 
ingredient design, 
formulation development, 
fermentation and production, 
texture analysis, 
sensory properties, 
manufacturing, and 
recipe generation. 
The example of a plant-based burger illustrates the tight integration of 
food science, 
biotechnology,
materials science, 
behavior science, and 
artificial intelligence 
to guide the development of plant-based burgers 
from raw ingredients to commercial products.} 
\label{fig02}
\end{figure}

We focus on the post-harvest stages and
follow an AI-powered cycle through 
ingredient design, 
formulation development, 
fermentation and production, 
texture analysis, 
sensory properties,
manufacturing, and 
recipe generation (Fig. \ref{fig02}). 
In the following sections, we highlight how AI-powered approaches, from generative models to digital twins, drive innovation across these steps.
%%%%%%%%%%%%%%%%%%%%%%%%%%%%%%%%%%%%%%%%%%%%%%%%%%%%%%%%%%%%%%%%%%%%%%%%%%%%%
%%%%%%%%%%%%%%%%%%%%%%%%%%%%%%%%%%%%%%%%%%%%%%%%%%%%%%%%%%%%%%%%%%%%%%%%%%%%%%
%%%%%%%%%%%%%%%%%%%%%%%%%%%%%%%%%%%%%%%%%%%%%%%%%%%%%%%%%%%%%%%%%%%%%%%%%%%%%
\section*{\sffamily{\bfseries{Results}}} 
%%%%%%%%%%%%%%%%%%%%%%%%%%%%%%%%%%%%%%%%%%%%%%%%%%%%%%%%%%%%%%%%%%%%%%%%%%%%%
%%%%%%%%%%%%%%%%%%%%%%%%%%%%%%%%%%%%%%%%%%%%%%%%%%%%%%%%%%%%%%%%%%%%%%%%%%%%%
% lisa, giorgia, rodrigo
{\sffamily{\bfseries{Ingredients: Driving functionality-first design.}}}
%%%%%%%%%%%%%%%%%%%%%%%%%%%%%%%%%%%%%%%%%%%%%%%%%%%%%%%%%%%%%%%%%%%%%%%%%%%%%
Novel food ingredients must balance sustainability, nutrition, sensory appeal, and functionality. Functionality includes water-holding, emulsification, gelation, viscosity, and stability that shape processing and structure. Many ingredients aim to replicate animal-derived functions, such as foaming in egg replacers; yet, matching performance remains challenging. Structure–function relationships link molecular architecture to activity. In food, macromolecular features of proteins, lipids, and fibers encode functionality. This raises the question to which extent AI can bridge this scale gap and translate molecular features into techno-functional predictions for a functionality-first design–build–test–learn cycle.

%%%%%%%%%%%%%%%%%%%%%%%%%%%%%%%%%%%%%%%%%%%%%%%%%%%%%%%%%%%%%%%%%%%%%%%%%%%%%
The success of animal-free ingredients depends on functionality and the ability to meet formulation and consumer requirements \cite{kyriakopoulou19}. These traits arise from molecular features, environmental conditions such as pH and temperature, and interactions with fats, fibers, and solutes. Proteins, lipids, and fibers provide core functionality, with proteins most versatile. The most promising alternative proteins today are wheat gluten and soy; they can  form fibrous textures under extrusion and shear \cite{kyriakopoulou19}. New sources include legumes, oilseeds, grains, algae, and leaves \cite{fasolin21}. Processing strongly shapes functionality: dry fractionation preserves native structures, while wet fractionation produces isolates that control solubility and gelation \cite{day13}. Lipids and fibers add complementary features, from stabilizing interfaces to modulating viscosity and lubrication. Since plant heterogeneity limits predictability, developers modify ingredients chemically, physically, or enzymatically. The field is now shifting toward functionality-enriched fractions for more consistent, cost-effective performance \cite{fasolin21}. 

%%%%%%%%%%%%%%%%%%%%%%%%%%%%%%%%%%%%%%%%%%%%%%%%%%%%%%%%%%%%%%%%%%%%%%%%%%%%%
AI and informatics enable functionality-driven ingredient design by linking molecular and mesoscopic features to macroscopic outcomes under processing constraints. Classification models support discovery; for example, a QSAR model identified glycyrrhizin as a natural emulsifier \cite{liu25a}. Regression models predict properties such as pectin viscosity to reduce cost and accelerate formulation \cite{siejak24}. Machine learning can predict techno-functional traits across crops with ${\textsf{R}}^2 > 0.80$ \cite{liepiang23}, while AI models can capture plant protein rheology with ${\textsf{R}}^2 > 0.99$ \cite{yilmaz25}. Deep learning increasingly links functionality to sensory design \cite{kraessig25}. A recent study optimized starch-based gels for saltiness and strength using e-tongue and oral-processing data \cite{meng25}. Multi-objective optimization can predict solubility, gel strength, and emulsification towards scalable formulation strategies \cite{liepiang23}. Together, these advances support closed-loop design–build–test–learn cycles. High-throughput screening, active learning, and digital twins operationalize models, while generative approaches target multiple criteria and bridge lab-to-plate gaps.

\begin{figure*}[t]
    \centering
   \includegraphics[width = 1.0\textwidth]{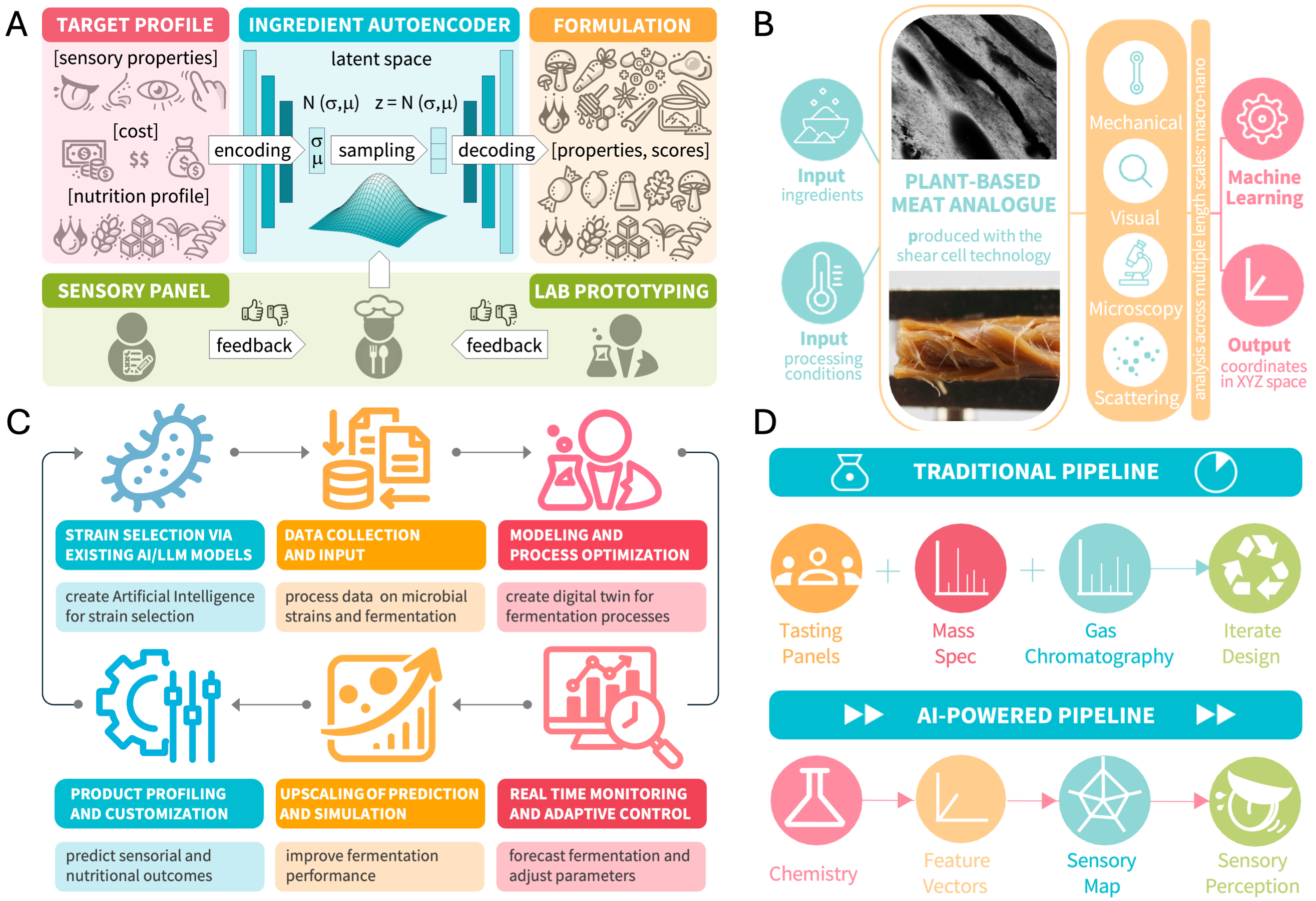}
    \caption{\sffamily{\bfseries{AI for food.}} 
(A) AI for Formulations. An iterative workflow to discover new formulations encodes target profiles such as sensory properties, cost, and nutritional profiles; generates formulation sets with deep learning; predicts properties and scores; and integrates laboratory prototyping and sensory panels in continuous feedback loops.
(B) AI for Texture.
Machine learning integrates multimodal data, from ingredients to process conditions, to optimize plant-based meat analogues, from microstructural appearance to macrostructural texture. Future machine learning tools could help reverse engineer ingredients and process conditions to achieve a desired appearance and texture.
(C) AI for Fermentation and Production. 
Artificial intelligence and large language models can accelerate the fermentation and production cycle during microbial strain selection, data collection, process optimization, real time monitoring and adaptive control, upscaling of prediction and simulation, and product profiling and customization.
(D) AI for Sensory Properties.
Optimizing the sensory perception of food is traditionally a costly, time-consuming process involving expert tasting panels and instrumental analysis, including mass spectrometry and gas chromatography. In contrast, an AI-powered pipeline has the potential to predict sensory perceptions from only the chemical or molecular make-up of food, eliminating the need for costly human panels and reducing the time to iterate product formulations.}   
\label{fig03}
\end{figure*}

%%%%%%%%%%%%%%%%%%%%%%%%%%%%%%%%%%%%%%%%%%%%%%%%%%%%%%%%%%%%%%%%%%%%%%%%%%%%%
Ingredient design still struggles with consistent performance across food matrices and key challenges remain. First, models often fail to generalize across ingredients and conditions. Second, nonlinear interactions in multicomponent systems complicate prediction and require hybrid physics-informed approaches. Third, sensory preferences vary across cultures, which demands diverse datasets. Fourth, models must link molecular features such as sequence, charge, and hydrophobicity to functionality in a transparent and scalable way. Addressing these challenges requires open, collaborative tools and an AI-driven ecosystem that predicts techno-functional performance from molecular features.\\[8.pt]
%%%%%%%%%%%%%%%%%%%%%%%%%%%%%%%%%%%%%%%%%%%%%%%%%%%%%%%%%%%%%%%%%%%%%%%%%%%%%
% karim
%%%%%%%%%%%%%%%%%%%%%%%%%%%%%%%%%%%%%%%%%%%%%%%%%%%%%%%%%%%%%%%%%%%%%%%%%%%%%
%%%%%%%%%%%%%%%%%%%%%%%%%%%%%%%%%%%%%%%%%%%%%%%%%%%%%%%%%%%%%%%%%%%%%%%%%%%%%
\noindent
{\sffamily{\bfseries{Formulations: Mapping flavor and recipe spaces.}}}
%%%%%%%%%%%%%%%%%%%%%%%%%%%%%%%%%%%%%%%%%%%%%%%%%%%%%%%%%%%%%%%%%%%%%%%%%%%%%
Ingredient functionality sets the foundation, but effective food design depends on how those ingredients interact in complex systems; formulation brings these components together. Meeting the dual mandate of sustainability and sensory fidelity requires moving beyond slow trial-and-error formulation development. AI offers a compelling approach to uncover hidden patterns that connect ingredients, formulations, production, texture, and sensory properties, and propose candidate formulations that are predicted to meet nutritional, functional, or sensorial targets \cite{notco21a,notco21b,notco21c}. Today, AI practitioners can draw on a broad collection of public resources including Recipe1M+ \cite{martin21}, FlavorGraph \cite{park21}, FlavorDB \cite{garg18}, USDA FoodData Central, and FAO/INFOODS, as well as domain-specific databases such as Phenol-Explorer, LipidBank, and the Volatile Compounds in Food database. Inputs include molecular descriptors, nutritional and functional properties, recipe structures, and process sequences; outputs are complete candidate formulations aligned with target criteria. AI-powered approaches combine latent-space learning through autoencoders with conditional generation, constraint-aware optimization, and expert-in-the-loop refinement (Fig. \ref{fig03}A). But can they truly capture the complex, non-linear interactions between ingredients, processes, and perception--even under sparse and noisy data--and unlock a new generation of food formulations?
%%%%%%%%%%%%%%%%%%%%%%%%%%%%%%%%%%%%%%%%%%%%%%%%%%%%%%%%%%%%%%%%%%%%%%%%%%%%%
%%%%%%%%%%%%%%%%%%%%%%%%%%%%%%%%%%%%%%%%%%%%%%%%%%%%%%%%%%%%%%%%%%%%%%%%%%%%%

Computational gastronomy has long shown that recipes and flavor chemistry form structured networks that we can mine for predictive patterns. The Flavor Network formalized ingredient compatibility via shared volatiles \cite{ahn11}, while FlavorGraph integrated one million recipes with $\sim$1,500 flavor molecules to generate graph embeddings for pairing prediction and ingredient clustering \cite{park21}. These studies have successfully shown that high-dimensional, multimodal food representations can capture the complexity of culinary design. In alternative proteins, industrial disclosures from NotCo patents \cite{notco21a,notco21b,notco21c} illustrate hybrid generative optimization pipelines. These systems embed ingredients and recipes in latent spaces using autoencoders and recurrent networks, condition sampling on feature vectors of target product profiles, refine candidates with constraint solvers such as sparse regression, and iteratively incorporate expert sensory feedback. This architecture mirrors multimodal generative trends in other fields and enables analogue discovery. It can identify non-obvious ingredient substitutions, while maintaining process feasibility. Other companies, including Climax Foods and Eat Just, deploy similar proprietary approaches, while academic research continues to build on Recipe1M+ \cite{martin21}, FlavorGraph \cite{park21}, and FlavorDB \cite{garg18}. Together, these efforts highlight a state of the art that blends representation learning with constraint-aware generation to accelerate formulation discovery.

Generative AI expands opportunities across protein diversification. For example, it can design plant-based cheeses that melt and stretch by identifying proteins that act like casein, the key driver of structure and meltability in animal cheese. It can discover ingredient blends that replicate meat textures and pinpoint volatile combinations that evoke familiar flavors \cite{tac26}. And it can improve for plant-based meats from sensory panel feedback and match expert predictions of panel rankings \cite{vandenbedem26}. Large-scale screening lets scientists focus on a few promising prototypes instead of hundreds of trials, which shortens the development cycle. In the near future, these tools will be able to tailor formulations to individual nutritional needs and taste preferences and enable personalized food design.

But significant challenges remain before AI can transform formulation discovery at scale. High-quality sensory datasets are limited and often fail to generalize across different food types. Even with detailed chemical and nutritional data, predicting human perception remains an open problem. Modeling ingredient performance under varying heat, shear, or storage conditions is incomplete. Cost, supply chain, allergen control, and clean-label requirements must be built into models as hard constraints. Trust among regulators, industry, and consumers will depend on transparent, auditable design processes. The near-term path will be hybrid, where AI proposes and humans refine, toward transparent formulation platforms that integrate constraints and predict performance at scale.\\[8.pt]
%%%%%%%%%%%%%%%%%%%%%%%%%%%%%%%%%%%%%%%%%%%%%%%%%%%%%%%%%%%%%%%%%%%%%%%%%%%%%
% yvonne
%%%%%%%%%%%%%%%%%%%%%%%%%%%%%%%%%%%%%%%%%%%%%%%%%%%%%%%%%%%%%%%%%%%%%%%%%%%%%
{\sffamily{\bfseries{Fermentation: Enabling smart bioreactors.}}}
%%%%%%%%%%%%%%%%%%%%%%%%%%%%%%%%%%%%%%%%%%%%%%%%%%%%%%%%%%%%%%%%%%%%%%%%%%%%%
Molecular and formulation design define the building blocks, but realizing performance requires precise control at the cellular and microbial scale. Precision fermentation---where engineered microorganisms are programmed to produce functional proteins such as casein or heme---has become central to protein diversification and sustainable food production. Its success depends on tightly coupled variables, strain stability, metabolic fluxes, nutrient composition, and bioreactor dynamics. AI has the potential to fuse these heterogeneous data with mechanistic insight to enable predictive, scalable, and cost-effective fermentation and production to address critical bottlenecks such as yield optimization and resource efficiency (Fig. \ref{fig03}C).

%%%%%%%%%%%%%%%%%%%%%%%%%%%%%%%%%%%%%%%%%%%%%%%%%%%%%%%%%%%%%%%%%%%%%%%%%%%%%
Optimizing fermentation requires a deep understanding of bioprocesses to maximize yield and efficiency while maintaining homogeneity, sterility, and scalability \cite{eastham24}. Kinetic bioprocess models remain widely used but are constrained by steady-state assumptions, and metabolic flux models for engineered strains often face computational bottlenecks. Experimental optimization through orthogonal design \cite{feng20} and response surface methodology \cite{sharma21} remains the industry standard, while multiscale feature engineering is still limited \cite{liao22}. 
Machine learning has proven valuable for modeling nonlinear fermentation dynamics \cite{liu09} and for real-time monitoring of anaerobic digestion \cite{jia22}. For example, deep learning soft sensors that combine convolutional neural networks with bidirectional long short-term memory architectures have achieved ${\textsf{R}}^2$ values up to 0.98 in predicting key anaerobic digestion state variables from real-time pH, electrical conductivity, and oxidation–reduction potential measurements \cite{jia22}. While AI and reinforcement learning have been successfully applied to sustainable protein fermentation \cite{cheng23} and co-culture optimization \cite{treloar20}, large-scale adoption of AI, and especially generative AI, remains limited.

%%%%%%%%%%%%%%%%%%%%%%%%%%%%%%%%%%%%%%%%%%%%%%%%%%%%%%%%%%%%%%%%%%%%%%%%%%%%%
AI can shift fermentation and production from reactive optimization to predictive management, with customized strains and self-optimizing processes. Machine learning models can optimize strain design, nutrient feeds, and bioreactor control, and deep learning approaches can capture non-linear interactions better than traditional regression \cite{singhal18}. Automation and AI-enhanced imaging \cite{austerjost21}, soft sensors \cite{butean25}, and spectroscopy \cite{cheng23} can improve monitoring and adaptive control. Algorithms such as support vector machines \cite{alejo18} and fuzzy inference systems \cite{akinade19} support robust predictions, while evolutionary optimization \cite{peng13} aids process tuning. Hybrid mechanistic-AI models \cite{sharma25} and explainable AI frameworks \cite{holzinger23} can improve transparency and decision-making. Training LLMs on unstructured bioprocess data--including text, images, and sensor logs--could support knowledge retrieval, troubleshooting, and experimental design and bridge biology, chemistry, engineering, and design through actionable insights, like predicting protein structures or cell morphology \cite{ghafarollahi24}. 

%%%%%%%%%%%%%%%%%%%%%%%%%%%%%%%%%%%%%%%%%%%%%%%%%%%%%%%%%%%%%%%%%%%%%%%%%%%%%
To realize this potential, the field must overcome the key barriers of data scarcity, data integration, and model transferability. Generating large experimental datasets remains labor intensive, but scaled-down bioreactors and small-data algorithms now offer new possibilities. At the same time, robust biosecurity and ethical frameworks will be critical to safely employ generative AI. Integration with genome-scale metabolic models and digital twins of bioreactor performance could create predictive platforms for design and scale-up. The collective goal is an integrated fermentation platform that links strain design, bioreactor dynamics, and real-time sensing to support predictive scale-up and reliable process optimization. \\[8.pt]
%%%%%%%%%%%%%%%%%%%%%%%%%%%%%%%%%%%%%%%%%%%%%%%%%%%%%%%%%%%%%%%%%%%%%%%%%%%%%
% miek
%%%%%%%%%%%%%%%%%%%%%%%%%%%%%%%%%%%%%%%%%%%%%%%%%%%%%%%%%%%%%%%%%%%%%%%%%%%%%%
%%%%%%%%%%%%%%%%%%%%%%%%%%%%%%%%%%%%%%%%%%%%%%%%%%%%%%%%%%%%%%%%%%%%%%%%%%%%%
{\sffamily{\bfseries{Texture: Learning structure–perception maps.}}}
%%%%%%%%%%%%%%%%%%%%%%%%%%%%%%%%%%%%%%%%%%%%%%%%%%%%%%%%%%%%%%%%%%%%%%%%%%%%%
Texture is a key quality attribute that shapes consumer acceptance. It emerges from coupled physical, mechanical, and sensory properties across scales—from molecular structure to oral processing—which complicates quantification and prediction. We distinguish between physical texture, measured through mechanical and rheological tests, and sensory texture, perceived through attributes such as chewiness or fibrousness. This distinction links measurable material behavior to human perception and defines how AI can connect mechanics with sensory response. Current approaches rely on costly sensory panels and labor-intensive tests such as cutting, tension, compression, shear, and rheology. The core challenge is to map intuitive descriptors like crispiness or chewiness onto measurable parameters that guide product design (Fig. \ref{fig03}B). This raises the question of how AI can integrate multimodal datasets to predict and engineer consumer-preferred textures and accelerate the adoption of sustainable proteins.

%%%%%%%%%%%%%%%%%%%%%%%%%%%%%%%%%%%%%%%%%%%%%%%%%%%%%%%%%%%%%%%%%%%%%%%%%%%%%
Every ingredient and product possesses a characteristic texture, which we can engineer through processing techniques such as extrusion, shear cell structuring, or 3D printing to mimic animal-based foods. Recent advances show the promise of machine learning for decoding texture–structure relationships: 
Constitutive neural networks can map mechanical signatures of plant-based meat alternatives onto interpretable and generalizable physics-based models \cite{stpierre23} that achieve ${\textsf{R}}^2$ values from 0.92 to 0.99 for commercially available plant-based meats \cite{stpierre24}.
Autoencoders can predict sensory texture attributes from rheological data and advance data analysis beyond traditional correlation methods \cite{kraessig25}. 
Classifiers such as support vector machines and neural networks can  identify snack freshness from mechanical and acoustic signals with up to 92\% accuracy \cite{sanahuja18}. 
Researchers are using machine learning to predict viscosity, gelation, and foaming of plant proteins \cite{dahl25,liepiang23}. 
Meanwhile, texture analysis from images is rapidly growing.
Transfer learning with models like VGGNet, ResNet, and DenseNet 
allows us to predict the cooking times for fish 
with potential extension to plant-based foods \cite{zhang25}, 
while ML-driven microscopic image analysis 
can help us identify droplets in colloidal systems 
to link microstructure to texture \cite{saalbrink25}. Beyond academic applications, commercial platforms already harness AI for texture optimization: For example, NotCo \cite{notco21a,notco21b,notco21c} and Climax Foods successfully deploy AI to mine ingredient databases, combine plant ingredients, and design formulations that replicate animal textures. 

%%%%%%%%%%%%%%%%%%%%%%%%%%%%%%%%%%%%%%%%%%%%%%%%%%%%%%%%%%%%%%%%%%%%%%%%%%%%%
AI accelerates prototyping by predicting textural outcomes before physical trials and  reduces time- and resource-intense trial-and-error cycles. In addition, ingredient design and process optimization can balance target textures with environmental impact, for example, to reduce greenhouse gas emissions. Another important application is personalization, for example designing softer foods for elderly populations to improve swallowing safety. LLMs can also convert consumer feedback such as too chewy or not crispy enough 
\cite{vandenbedem26} into quantifiable parameters for product adjustment,  and bridge the gap between perception and engineering \cite{dunne25}. To address data sparsity, LLMs can generate synthetic training data by pairing realistic texture descriptors with mechanical properties. Ideally, AI can integrate data across multiple length scales, from molecular interactions to macroscopic structure, to capture the hierarchical origins of texture. Inversely, future machine learning tools could help reverse engineer ingredients and process conditions to achieve a desired texture. More broadly, generative design tools, similar to those used for inorganic materials \cite{zeni25}, hold the potential to propose ingredient–process combinations that meet desired textural constraints.

%%%%%%%%%%%%%%%%%%%%%%%%%%%%%%%%%%%%%%%%%%%%%%%%%%%%%%%%%%%%%%%%%%%%%%%%%%%%%
Significant obstacles remain before AI-driven texture engineering will truly become routine. The biggest challenge is the lack of large, open datasets that couple textural, sensory, and processing data. Industrial data sharing is rare and data fusion remains technically challenging, as mechanical, rheological, visual, and compositional data are inherently multimodal and often unstructured. The ultimate vision would be an AI system for inverse design--with a specific target texture as input and optimal ingredients and processes as output--that transforms texture development into a predictive, model-guided science. \\[8.pt]
%%%%%%%%%%%%%%%%%%%%%%%%%%%%%%%%%%%%%%%%%%%%%%%%%%%%%%%%%%%%%%%%%%%%%%%%%%%%%
% lisa, giorgia, rodrigo
%%%%%%%%%%%%%%%%%%%%%%%%%%%%%%%%%%%%%%%%%%%%%%%%%%%%%%%%%%%%%%%%%%%%%%%%%%%%%
%%%%%%%%%%%%%%%%%%%%%%%%%%%%%%%%%%%%%%%%%%%%%%%%%%%%%%%%%%%%%%%%%%%%%%%%%%%%%
{\sffamily{\bfseries{Sensory: Learning perceptual representations.}}}
%%%%%%%%%%%%%%%%%%%%%%%%%%%%%%%%%%%%%%%%%%%%%%%%%%%%%%%%%%%%%%%%%%%%%%%%%%%%%
Ultimately, food is not a material design challenge, but a sensory experience. Aroma, flavor, and texture are the strongest determinants of consumer acceptance. The central challenge is to map measurable physical and chemical properties of ingredients to the multidimensional space of human perception. Machine learning provides a framework to learn this mapping and enables both the prediction of sensory profiles and the computational design of novel ingredients with desired characteristics (Fig. \ref{fig03}D). But how far are current AI models from accurately predicting human sensory perception and from generating novel flavor and aroma experiences?

%%%%%%%%%%%%%%%%%%%%%%%%%%%%%%%%%%%%%%%%%%%%%%%%%%%%%%%%%%%%%%%%%%%%%%%%%%%%%
Traditional sensory science has relied on human panels for perceptual evaluation and on analytical instruments such as gas chromatography–mass spectrometry for chemical composition. Panels remain the gold standard, but are slow and expensive, while instruments offer precision without perceptual insight. Early machine learning approaches combined biomimetic sensors, such as E-noses, with classifiers \cite{munekata23}, but required heavy feature engineering and were limited to narrow classification tasks \cite{karakaya19}. The field has since shifted toward deep representation learning \cite{bengio12}, which learns general-purpose sensory maps directly from data. 
The Principal Odor Map exemplifies this trend: a graph neural network organizes molecules into a vector space where distance encodes perceptual similarity with human-level accuracy in odor labeling: 
On a prospective validation set of 323 novel odorants, the model's predictions closely matched a trained panel mean with per-molecule correlations approaching {\textsf{R}} $\approx$ 0.6 in representative cases \cite{lee23}.
Researchers now employ attention-based architectures to extend this mapping to molecular mixtures \cite{tom25}. Algorithmic tools for predictive sensory modeling are maturing rapidly, but progress is constrained by the scarcity and cost of high-quality sensory data.

%%%%%%%%%%%%%%%%%%%%%%%%%%%%%%%%%%%%%%%%%%%%%%%%%%%%%%%%%%%%%%%%%%%%%%%%%%%%%
Learned sensory representations open powerful opportunities for food innovation. High-throughput virtual screening of molecular libraries can identify novel flavor compounds for plant-based meats and accelerate product development \cite{kuhl25}. Integrating predictive models with low-cost, miniaturized sensors enables real-time quality monitoring in manufacturing and packaging \cite{istif23}, for example with wireless freshness indicators that reduce food waste \cite{watson21}. These tools also provide systematic frameworks for biomimicry to predict how modifications of ingredients or processes alter the final sensory profile. Still, a feedback loop remains essential: computational predictions must be continuously validated and refined with real-world experiments to ensure consumer relevance. 

%%%%%%%%%%%%%%%%%%%%%%%%%%%%%%%%%%%%%%%%%%%%%%%%%%%%%%%%%%%%%%%%%%%%%%%%%%%%%
Today, the bottleneck is no longer algorithmic capacity but high-quality data acquisition \cite{rombach22}. Deep learning models require large cross-scale datasets, yet sensory panels cannot generate data at scale to capture the vast chemical space of food \cite{vandenbedem26}. The path forward lies in closing the loop between modeling and experiment. Generative models such as transformers and diffusion architectures are algorithmically ready, but their success depends on new data pipelines. Developing robust, low-cost hardware and embedding it into automated platforms will be critical to realize the full potential of AI for sensory science. The ultimate vision is a self-driving sensory lab that integrates chemical synthesis, sensory characterization, and machine learning to autonomously generate, predict, and validate flavor, aroma, and texture perception across diverse food systems. \\[8.pt]
%%%%%%%%%%%%%%%%%%%%%%%%%%%%%%%%%%%%%%%%%%%%%%%%%%%%%%%%%%%%%%%%%%%%%%%%%%%%%
% nik
%%%%%%%%%%%%%%%%%%%%%%%%%%%%%%%%%%%%%%%%%%%%%%%%%%%%%%%%%%%%%%%%%%%%%%%%%%%%%
%%%%%%%%%%%%%%%%%%%%%%%%%%%%%%%%%%%%%%%%%%%%%%%%%%%%%%%%%%%%%%%%%%%%%%%%%%%%%
{\sffamily{\bfseries{Manufacturing: Bridging lab and industrial scale.}}}
%%%%%%%%%%%%%%%%%%%%%%%%%%%%%%%%%%%%%%%%%%%%%%%%%%%%%%%%%%%%%%%%%%%%%%%%%%%%%
Even the best foods will fail if they cannot be produced consistently and affordably. Manufacturing convincing meat analogues is inherently complex because technical goals such as texture, flavor, appearance, and shelf life interact non-linearly with environmental and economic constraints. Relevant datasets span composition, structure, rheology, thermal behavior, processing steps such as extrusion or fermentation, as well as telemetry, sensory, and consumer data. 
Through digital twins, computer vision, and other modes, AI is emerging as a key technology to integrate these multimodal datasets and streamline process optimization. Typical model inputs include formulation and process profiles, while outputs can range from texture, color, and flavor to cost and environmental impact.
But can these models deliver reliable, scalable, and cost-effective production paths without compromising sensory quality or sustainability goals?

%%%%%%%%%%%%%%%%%%%%%%%%%%%%%%%%%%%%%%%%%%%%%%%%%%%%%%%%%%%%%%%%%%%%%%%%%%%%%
AI and ML span across the manufacturing pipeline, from formulation and process design to in-line quality control. High-moisture extrusion produces fibrous structures that mimic muscle, but outcomes depend on moisture, temperature, and screw configuration. Bayesian optimization explores this space more efficiently than trial-and-error methods \cite{dinali24}. In a pilot study, it identified optimal parameters in ten trials versus fifteen for response-surface methodology and reduced validation errors from 61\% to below 15\% \cite{jiang25}. Digital twins support soft sensing and control, but most applications remain at pilot scale \cite{abdurrahman25}. Data fragmentation persists across proprietary logs and repositories with inconsistent metadata. To address these limitations, FAIR data principles \cite{top22} and LLM-based extraction \cite{bolucu25} have successfully employed knowledge graphs to link formulation, process, structure, and sensory performance.

%%%%%%%%%%%%%%%%%%%%%%%%%%%%%%%%%%%%%%%%%%%%%%%%%%%%%%%%%%%%%%%%%%%%%%%%%%%%%
AI is opening multiple opportunities for advancing alternative protein manufacturing. Engineers can use machine learning and Bayesian optimization to tune high-moisture extrusion parameters to target fibrousness, yield, and energy efficiency, while digital twins evaluate what-if simulations for scale-up and real-time sensing \cite{watson25,dinali24,goodfoodinstitute}. Food scientists can use informatics pipelines to rank plant proteins and design blends at scale using rheology–functionality maps that allow ingredient discovery well beyond commodity sources \cite{jiang25,top22}. LLMs further act as formulation copilots, transform unstructured texts into structured knowledge graphs, and suggest experiments to resolve uncertainty \cite{top22}. At the same time, computer vision and hyperspectral analytics can advance in-line quality assurance by linking color and structure to sensory proxies to enable continuous assessment during production \cite{goodfoodinstitute}. Together, these approaches accelerate prototyping, reduce waste, and bridge lab and industrial manufacturing.

%%%%%%%%%%%%%%%%%%%%%%%%%%%%%%%%%%%%%%%%%%%%%%%%%%%%%%%%%%%%%%%%%%%%%%%%%%%%%
Scaling these tools requires high-quality, shareable datasets that link formulations, processes, structures, and sensory outcomes with standardized metadata. Proprietary and fragmented data currently hinder model transfer and benchmarking and underscore the need for domain-specific FAIR implementations, shared testbeds, and community baselines \cite{abdurrahman25}. Ingredient variability remains a crucial bottleneck and robust hybrid models are needed to achieve reliable process control \cite{watson25,dinali24}. Industrial adoption also demands capital investment, workforce training, and validation on real production lines since most food digital twins have not been tested at scale \cite{goodfoodinstitute}. Shared data standards and LLM-enabled knowledge bases can bridge lab and industrial scale and enable consistent, cost-efficient manufacturing at or below price parity.\\[8.pt]
%%%%%%%%%%%%%%%%%%%%%%%%%%%%%%%%%%%%%%%%%%%%%%%%%%%%%%%%%%%%%%%%%%%%%%%%%%%%%
% kristina, anna, dan
%%%%%%%%%%%%%%%%%%%%%%%%%%%%%%%%%%%%%%%%%%%%%%%%%%%%%%%%%%%%%%%%%%%%%%%%%%%%%
%%%%%%%%%%%%%%%%%%%%%%%%%%%%%%%%%%%%%%%%%%%%%%%%%%%%%%%%%%%%%%%%%%%%%%%%%%%%%
{\sffamily{\bfseries{Recipes: Personalizing food experiences.}}}
%%%%%%%%%%%%%%%%%%%%%%%%%%%%%%%%%%%%%%%%%%%%%%%%%%%%%%%%%%%%%%%%%%%%%%%%%%%%%
Beyond production, food preparation shapes consumer experience. Individuals and foodservice operations must balance enjoyment with health, cost, convenience, and allergens while advancing broader goals such as human and planetary health. LLMs offer strong potential because they draw on large recipe datasets and integrate sources such as Recipes1M, USDA FoodData Central, the HESTIA climate database, and Food.com reviews. The key challenge is to embed human knowledge so that models generalize across diverse users, dietary needs, and cultural traditions.

%%%%%%%%%%%%%%%%%%%%%%%%%%%%%%%%%%%%%%%%%%%%%%%%%%%%%%%%%%%%%%%%%%%%%%%%%%%%%
Recent research evaluates and refines LLMs for recipe generation, revision, and preference prediction. Progress relies on fine-tuning, retrieval-augmented generation, and integration with optimization tools. Benchmarks on Recipe1M show that off-the-shelf models produce unreliable outputs \cite{mohbat24}. Fine-tuning improves performance \cite{li24} and supports revision tasks \cite{senath24}. A generative model trained on 2,216 curated burger recipes explored $8.9 \times 10^{43}$ candidate formulations and identified Pareto-efficient recipes that reduced environmental impact without compromising taste \cite{tac26}. In blinded restaurant surveys, some AI-generated burgers outperformed the BigMac benchmark in flavor, texture, and overall liking \cite{tac26a}. Other work demonstrates retrieval-augmented generation for recipes \cite{liu25b} and sustainable menu design \cite{thomas25}. Some LLMs now match expert food scientists in preference modeling \cite{thomas25}. Combined with optimization algorithms, these models can generate recipes that meet nutritional and sustainability targets.

%%%%%%%%%%%%%%%%%%%%%%%%%%%%%%%%%%%%%%%%%%%%%%%%%%%%%%%%%%%%%%%%%%%%%%%%%%%%%
The ability to generate recipes under explicit constraints creates opportunities across individual and institutional contexts. For individuals, fine-tuned LLMs could power AI health coaches that teach users how to modify meals to be more sustainable, nutritious, or culturally appropriate, while respecting restrictions such as allergies or religious requirements. Chefs and students could practice recipe generation in educational applications, receiving AI-driven feedback to refine their skills \cite{yang24}. In institutional settings, LLMs could support foodservice operations in offices, universities, or hospitals, tailoring  descriptions, ingredients, and preparation methods under strict constraints. This is particularly valuable in programs such as “Food as Medicine,” which deliver cost- and time-sensitive meals for patients with chronic conditions after hospital discharge \cite{rosas25}. In all these scenarios, AI could accelerate constrained recipe generation, broaden access to healthy and sustainable foods, and scale public health interventions that rely on tailored meal planning.

%%%%%%%%%%%%%%%%%%%%%%%%%%%%%%%%%%%%%%%%%%%%%%%%%%%%%%%%%%%%%%%%%%%%%%%%%%%%%
Despite this promise, important challenges remain. Off-the-shelf LLMs perform unevenly across cultures, fail to capture diverse preferences, and produce numerically inconsistent or unsafe outputs. They lack molecular-level understanding and have limited knowledge of food safety. Training models from scratch requires substantial resources. These limitations introduce risks of impractical or unsafe recipes and increase environmental cost. Mitigation strategies include integrating external datasets, using open-source pre-trained models, and coupling LLMs with optimization tools. Beyond technical concerns, systems must preserve user agency and support human expertise. A more human-centered approach would allow LLMs to assist with drafting and revision, while users adapt recipes to their needs, preferences, and cultures.
%%%%%%%%%%%%%%%%%%%%%%%%%%%%%%%%%%%%%%%%%%%%%%%%%%%%%%%%%%%%%%%%%%%%%%%%%%%%%
% markus and david
%%%%%%%%%%%%%%%%%%%%%%%%%%%%%%%%%%%%%%%%%%%%%%%%%%%%%%%%%%%%%%%%%%%%%%%%%%%%%
%%%%%%%%%%%%%%%%%%%%%%%%%%%%%%%%%%%%%%%%%%%%%%%%%%%%%%%%%%%%%%%%%%%%%%%%%%%%%
\section*{\sffamily{\bfseries{Conclusions}}}
%%%%%%%%%%%%%%%%%%%%%%%%%%%%%%%%%%%%%%%%%%%%%%%%%%%%%%%%%%%%%%%%%%%%%%%%%%%%%
AI is beginning to transform food science across the entire innovation cycle,
from 
ingredient design, 
formulation development, and
fermentation and production to 
texture analysis, 
sensory properties, 
manufacturing, and 
recipe generation (Fig. \ref{fig02}).
Yet, the field's true potential not only lies in optimizing today's processes, but also in charting a bold trajectory for the future of food as a programmable, predictive science. Building on recent advances, we envision four major areas that define the frontier of AI for sustainable food production:\\[6.pt]
%%%%%%%%%%%%%%%%%%%%%%%%%%%%%%%%%%%%%%%%%%%%%%%%%%%%%%%%%%%%%%%%%%%%%%%%%%%%%
{\sffamily{\bfseries{Future materials for food innovation.}}}
%%%%%%%%%%%%%%%%%%%%%%%%%%%%%%%%%%%%%%%%%%%%%%%%%%%%%%%%%%%%%%%%%%%%%%%%%%%%% 
Treating food as a programmable biomaterial creates opportunities to design ingredients from first principles: texture, functionality, stability, and sensory perception emerge from multiscale interactions among proteins, carbohydrates, lipids, and processing conditions \cite{gordon25}. This perspective enables the design of edible films and gels with controlled nutrient release and plant-based meat analogues with tailored mechanical and sensory performance. Advances in protein design, including AlphaFold and diffusion-based generative models \cite{abramson24}, point toward de novo edible proteins with tunable functionality. Early work in cellular agriculture, such as engineered fat and muscle cells with modified composition \cite{yuen23}, demonstrates this potential, while microbial and fungi-based fermentation offer complementary routes. Mycelium fermentation can upcycle agricultural side-streams such as brewers’ spent grain or fruit pomace into protein- and fiber-rich ingredients with distinctive textures and clear links to circularity \cite{vervenne25}. A key opportunity lies in integrating molecular design with tunable processing, including fermentation and upcycling, to build predictive platforms that generate new food materials with targeted properties and reduced waste.\\[6.pt]
%%%%%%%%%%%%%%%%%%%%%%%%%%%%%%%%%%%%%%%%%%%%%%%%%%%%%%%%%%%%%%%%%%%%%%%%%%%%%
{\sffamily{\bfseries{Scientific machine learning for food systems.}}}
%%%%%%%%%%%%%%%%%%%%%%%%%%%%%%%%%%%%%%%%%%%%%%%%%%%%%%%%%%%%%%%%%%%%%%%%%%%%%
Scientific Machine Learning (SciML) \cite{rackauckas20}
hardwires physical, chemical, and biological principles
directly into AI architectures
and provides a foundation 
for trustworthy and scalable food innovation.
Physics-Informed Neural Networks (PINNs) \cite{raissi19}
enforce conservation laws
in data-sparse settings \cite{karniadakis21}
such as fermentation and extrusion.
Constitutive Artificial Neural Networks (CANNs) \cite{linka23}
embed mechanical laws 
to discover physically interpretable models 
of food texture and rheology \cite{stpierre24}.
Sparse Identification of Nonlinear Dynamics (SINDy) \cite{brunton16}
discovers compact governing equations
from experimental data and 
can reveal the kinetics of ingredient interactions or microbial growth.
The opportunity ahead is to couple physics-based ML with data-driven models to achieve interpretable, generalizable predictions under sparse data and enable mechanism-aware food design.
%%%%%%%%%%%%%%%%%%%%%%%%%%%%%%%%%%%%%%%%%%%%%%%%%%%%%%%%%%%%%%%%%%%%%%%%%%%%%
\begin{figure*}[h]
    \centering
   \includegraphics[width = 0.72\textwidth]{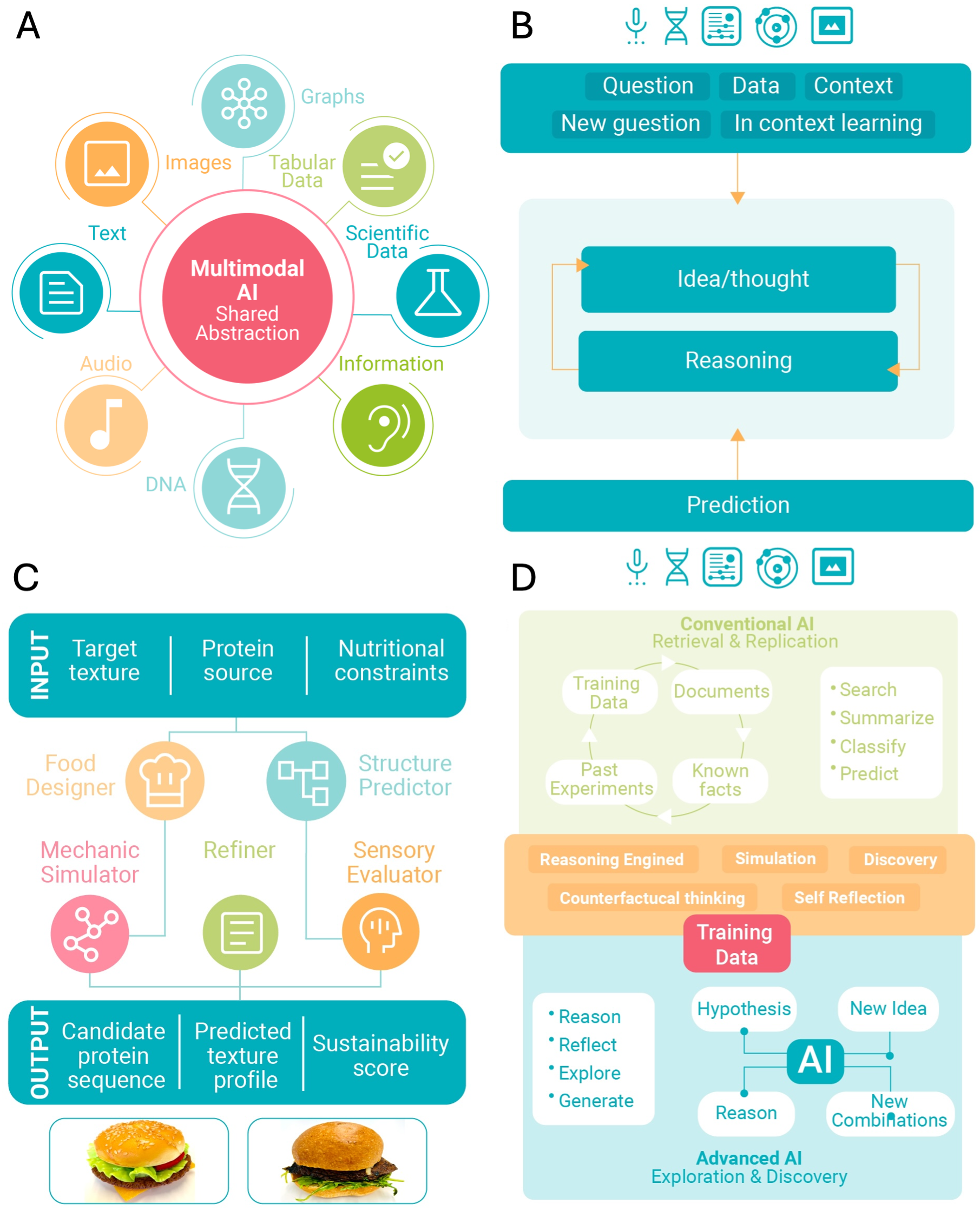}
    \caption{\sffamily{\bfseries{Next-generation AI systems for scientific discovery in food.}}
(A) Multimodal AI integrates diverse data types such as text, images, graphs, and molecular structures; 
(B) first-principles AI reasoning models perform iterative thinking and hypothesis generation; 
(C) a possible workflow applies these capabilities to edible protein design and predicts structure, mechanics, and sensory outcomes by explicitly modeling disparate concepts and their relationships; and 
(D) the transition from conventional retrieval-based AI to reasoning-enabled, self-reflective AI enables exploration, discovery, and ultimately, the generation of novel, functional foods.}    
\label{fig04}
\end{figure*} \\[6.pt]
%%%%%%%%%%%%%%%%%%%%%%%%%%%%%%%%%%%%%%%%%%%%%%%%%%%%%%%%%%%%%%%%%%%%%%%%%%%%%
%%%%%%%%%%%%%%%%%%%%%%%%%%%%%%%%%%%%%%%%%%%%%%%%%%%%%%%%%%%%%%%%%%%%%%%%%%%%%
{\sffamily{\bfseries{Automation and self-driving labs.}}}
%%%%%%%%%%%%%%%%%%%%%%%%%%%%%%%%%%%%%%%%%%%%%%%%%%%%%%%%%%%%%%%%%%%%%%%%%%%%%
Empirical trial-and-error methods based on rheology, sensory panels, and compositional analysis have long shaped food science, but they remain slow, costly, and hard to scale. Generative AI, robotics, and high-throughput experimentation enable self-driving labs that design, test, and refine food materials within a closed-loop design–build–test–learn cycle, accelerating systematic exploration across scales.
Recent advances in automated model discovery with custom-designed constitutive neural networks, show how AI can autonomously identify the best models, parameters, and even experimental protocols directly from data \cite{linka23}--a method already successfully applied to compare plant-based and animal meats \cite{stpierre24}, deli products \cite{stpierre25}, and fungi-based steak \cite{vervenne25}. 
Agentic AI systems such as Sparks \cite{ghafarollahi25}, BioDiscoveryAgent \cite{roohani24}, and the Virtual Lab \cite{swanson25}  demonstrate the capacity to generate hypotheses, perform simulations, and report results without human intervention. In food, such systems could integrate genome-scale metabolic models, digital twins of bioreactors, and in-line quality sensors \cite{buehler24,ghafarollahi24}. A fully automated discovery pipeline that links synthesis, characterization, and modeling would substantially accelerate innovation in sustainable, personalized, and scalable food production.\\[6.pt]
%%%%%%%%%%%%%%%%%%%%%%%%%%%%%%%%%%%%%%%%%%%%%%%%%%%%%%%%%%%%%%%%%%%%%%%%%%%%%
{\sffamily{\bfseries{Deep reasoning models.}}}
%%%%%%%%%%%%%%%%%%%%%%%%%%%%%%%%%%%%%%%%%%%%%%%%%%%%%%%%%%%%%%%%%%%%%%%%%%%%%
AI models with reasoning capabilities will become central to self-driving lab infrastructure (Fig. \ref{fig04}). The emerging discipline of AI for Food will benefit from rapid progress in reasoning models and agentic systems that design and execute tasks with external tools \cite{boiko23}. Current generative models propose molecules and recipes, but they rely on static datasets and struggle to assess feasibility, manufacturability, or emergent properties. Closing this gap requires systems that generate protein sequences, simulate folding and mechanics, and predict sensory outcomes such as flavor and texture within an interpretable, self-improving loop. Deep reasoning models define the next frontier beyond pattern recognition, and instead hypothesize, simulate, and generate data. They integrate protein sequences, structural mechanics, metabolic fluxes, and sensory responses to predict how molecular composition translates into taste, texture, and stability \cite{gordon25}. They can reduce the simulation-to-reality gap by validating predictions against experiments and updating models in real time. Such approaches can guide experiment selection \cite{stpierre23}, integrate results into finite element simulations of oral processing \cite{peirlinck24}, and predict sensory perception from physical principles \cite{stpierre24}. By embedding sustainability constraints, cultural diversity, and personalized health data, these systems can design foods that meet functional, nutritional, and societal goals. The ultimate goal is a closed-loop, reasoning-driven platform that designs, tests, and refines food materials with predictive accuracy across scales.\\[8.pt]
%%%%%%%%%%%%%%%%%%%%%%%%%%%%%%%%%%%%%%%%%%%%%%%%%%%%%%%%%%%%%%%%%%%%%%%%%%%%%
{\sffamily{\bfseries{Limitations and risks.}}}
%%%%%%%%%%%%%%%%%%%%%%%%%%%%%%%%%%%%%%%%%%%%%%%%%%%%%%%%%%%%%%%%%%%%%%%%%%%%%
This review uses sustainable proteins as a testbed for AI-enabled food systems and focuses on post-harvest product development, from ingredients to recipes. It does not cover upstream agriculture, supply chains, packaging, safety testing, or retail analytics, domains that also adopt AI and merit dedicated reviews. Benchmarking remains limited by heterogeneous datasets, proprietary silos, and weak metadata standards, which constrain reproducibility and global representation. AI integration also introduces technical and societal risks, including limited transferability, domain shift, and bias from sparse data. In food systems, these risks carry direct safety implications: inconsistent outputs may yield unsafe formulations, opaque optimization may hide allergen or nutritional trade-offs, and automated processes raise safety and biosecurity concerns. Responsible use of AI requires transparent benchmarks, physics-informed constraints, human-in-the-loop validation, regulatory alignment, and inclusive data design. AI must augment expert oversight to ensure safe, scalable, and sustainable innovation.
\\[6.pt]
%%%%%%%%%%%%%%%%%%%%%%%%%%%%%%%%%%%%%%%%%%%%%%%%%%%%%%%%%%%%%%%%%%%%%%%%%%%%%
{\sffamily{\bfseries{Outlook.}}}
%%%%%%%%%%%%%%%%%%%%%%%%%%%%%%%%%%%%%%%%%%%%%%%%%%%%%%%%%%%%%%%%%%%%%%%%%%%%%
Taken together, these four areas define a path toward a predictive and design-driven science of food. AI can link molecular composition to structure, processing, and perception and integrate these levels into closed-loop systems that design, test, and refine foods across scales. From programmable biomaterials and physics-informed models to self-driving laboratories and reasoning-based AI, the field now has the foundations to move from empirical optimization to mechanism-aware design. Realizing this vision will require interoperable data standards, rigorous benchmarking, and platforms that integrate experiments, simulations, and learning. Progress will also depend on human-centered design that embeds cultural context, nutrition, and safety into AI systems. If these challenges are addressed, AI will not only accelerate food innovation, but establish a new paradigm in which food design becomes predictive, scalable, and aligned with both human and planetary health.\\
%%%%%%%%%%%%%%%%%%%%%%%%%%%%%%%%%%%%%%%%%%%%%%%%%%%%%%%%%%%%%%%%%%%%%%%%%%%%%
\backmatter
%%%%%%%%%%%%%%%%%%%%%%%%%%%%%%%%%%%%%%%%%%%%%%%%%%%%%%%%%%%%%%%%%%%%%%%%%%%%%
\bmhead*{\sffamily{\bfseries{Acknowledgments}}}
%%%%%%%%%%%%%%%%%%%%%%%%%%%%%%%%%%%%%%%%%%%%%%%%%%%%%%%%%%%%%%%%%%%%%%%%%%%%%
We thank 
Dr. Heideh Fattaey for stimulating discussions on the future of food,
Ryan Yow for helping to design Figure \ref{fig03}C, and
Goran Atanasovski for illustrating Figures \ref{fig03} to \ref{fig04}. 
%Singapore Institute of Food \& Biotechnology Innovation
%
This project was supported 
by 
the USDA Grant FA9550-23-1-0606 to DLK,
by 
the BBSRC Grant BB/Y008510/1, 
the ERC Grant DEUSBIO-949080, and 
the Bezos Earth Fund through the Bezos Centre for Sustainable Protein BCSP/IC/001 to RLA,
by Novo Nordisk Foundation grant NNF23OC0085919 to MS,
by the NSF Graduate Research Fellowship to SRSP, 
by the UK National Alternative Protein Innovation Centre NAPIC, an Innovation and Knowledge Centre funded by the Biotechnology and Biological Sciences Research Council BBSRC and Innovate UK Grant BB/Z516119/1 to NW,
%
%[... send us your acknowledgements to be added here ...]
%
and by 
the Food@Stanford Snack Grant, 
the Stanford SDSS Accelerator Grant, 
the NSF CMMI Award 2320933, and
the ERC Advanced Grant 101141626 to EK.
%%%%%%%%%%%%%%%%%%%%%%%%%%%%%%%%%%%%%%%%%%%%%%%%%%%%%%%%%%%%%%%%%%%%%%%%%%%%%
\bmhead*{\sffamily{\bfseries{Conflicts of interest}}} 
The authors declare no competing interests.
%%%%%%%%%%%%%%%%%%%%%%%%%%%%%%%%%%%%%%%%%%%%%%%%%%%%%%%%%%%%%%%%%%%%%%%%%%%%%
\bmhead*{\sffamily{\bfseries{Author information}}} 
Bianca Datta, Good Food Institute, Washington, USA;
Markus J. Buehler, Massachusetts Institute of Technology, Cambridge, USA;
Yvonne Chow; Singapore Institute of Food \& Biotechnology Innovation SIFBI and Agency for Science, Technology and Research A*STAR, Nanos, Singapore;
Kristina Gligoric, Johns Hopkins University, Baltimore, USA;
Dan Jurafsky, Stanford University, Stanford, USA;
David L. Kaplan, Tufts University, Medford, USA;
Rodrigo Ledesma-Amaro,
Giorgia Del~Missier,
Lisa Neidhardt; Bezos Centre for Sustainable Protein; Microbial Food Hub; 
Centre for Engineering Biology, Imperial College London, London, UK;
Karim Pichara, NotCo, San Francisco, USA;
Benjamin Sanchez-Lengeling, University of Toronto, Toronto, Canada;
Miek Schlangen, University of Southern Denmark, Odense, Denmark;
Skyler R. St. Pierre, Stanford University, Stanford, USA;
Ilias Tagkopoulos, USDA/NIFA AI Institute for Next-Generation Food Systems,
University of California, Davis, USA;
Anna Thomas, Stanford University, Stanford, USA;
Nik Watson, University of Leeds, Leeds, UK and
National Alternative Protein Innovation Centre NAPIC, UK;
Ellen Kuhl, Stanford University, Stanford, USA
%%%%%%%%%%%%%%%%%%%%%%%%%%%%%%%%%%%%%%%%%%%%%%%%%%%%%%%%%%%%%%%%%%%%%%%%%%%%%
%%%%%%%%%%%%%%%%%%%%%%%%%%%%%%%%%%%%%%%%%%%%%%%%%%%%%%%%%%%%%%%%%%%%%%%%%%%%%
\bmhead*{\sffamily{\bfseries{Authors' contributions}}} 
BD and EK designed the layout, 
BD, MJB, YC, KG, DJ, DLK, RLA, GDM, LN, KP, BSL, MS, SRSP, IT, AT, NW, and EK wrote and edited the paper. 
%%%%%%%%%%%%%%%%%%%%%%%%%%%%%%%%%%%%%%%%%%%%%%%%%%%%%%%%%%%%%%%%%%%%%%%%%%%%%
%%%%%%%%%%%%%%%%%%%%%%%%%%%%%%%%%%%%%%%%%%%%%%%%%%%%%%%%%%%%%%%%%%%%%%%%%%%%%
%%%%%%%%%%%%%%%%%%%%%%%%%%%%%%%%%%%%%%%%%%%%%%%%%%%%%%%%%%%%%%%%%%%%%%%%%%%%%

\end{document}